# Heterochromatic nonlinear optical responses in upconversion nanoparticles for point spread function engineering

Chaohao Chen[1], Baolei Liu[1], Yongtao Liu[1], Jiayan Liao[1], Xuchen Shan[1], Fan Wang[1,2]*, and Dayong Jin[1,3]*

[1]Institute for Biomedical Materials & Devices (IBMD), Faculty of Science, University of Technology Sydney, NSW 2007, Australia

[2]School of Electrical and Data Engineering, Faculty of Engineering and Information Technology, University of Technology Sydney, Ultimo 2007, Australia

[3]UTS-SUStech Joint Research Centre for Biomedical Materials & Devices, Department of Biomedical Engineering, Southern University of Science and Technology, Shenzhen, China

*Correspondence to fan.wang@uts.edu.au, dayong.jin@uts.edu.au

**Abstract**

Point spread function (PSF) engineering of the emitter can code higher spatial frequency information of an image to break diffraction limit but suffer from the complexed optical systems. Here we present a robust strategy to simultaneously achieve diverse PSFs from upconversion nanoparticles under a single doughnut-shape scanning beam. By saturating the four-photon state, the high-frequency information can be extracted through the doughnut emission PSF. In contrast, the complementary lower frequency information can be carried out by the Gaussian-like emission PSF, as a result of over-saturated at the two-photon state. With the Fourier domain heterochromatic fusion, we verify the capability of the synthesised PSF to cover both low and high-frequency information, yielding the overall enhanced image quality. We show a spatial resolution of 40 nm, 1/24$^{th}$ of the excitation wavelength. This work suggests a new scope for developing nonlinear multi-colour emitting probes to improve image quality and noise control in nanoscopy.

**Introduction**

To overcome the optical diffraction limit, techniques using patterned light excitation, such as fringes (*1*), (*2*) and speckles (*3*), (*4*) become the essential scheme for super-resolution imaging. Approaches based on the point-scanning nonlinear illumination (*5*) can obtain extraordinary high efficacy in resolution enhancement. Stimulated emission depletion (STED) microscopy is a prominent example of point spread functions (PSF) engineering by generating doughnut-shaped beam (*6*) to resolve higher spatial frequencies in the doughnut spot than a diffraction-limited Gaussian PSF (*7*). By using exclusively doughnut-shaped excitation spots to saturate fluorescence, new modalities (*8*), (*9*), (*10*), (*11*) have been demonstrated to extract the extraordinary information at the higher spatial frequency for resolution enhancement. In the latest development, using upconversion nanoparticles (UCNPs) (*12*), single scan using two synchronised laser beams has resulted in a fluorescence emission difference (FED) microscopy approach that uses the two-colour PSF engineering. Explorations of the nonlinear properties of fluorescent probes, either inducing binary stochastic methods (*13*), (*14*), quantum coherent control (*15*) or under saturated conditions (*16*), (*17*), (*18*), (*19*), have resulted in the new approaches of stimulated emission (*6*), (*20*), ground state depletion (*9*), absorption (*21*), (*22*) or fluorescence photo-switching (*23*). Nevertheless, as a prerequisite to extract the broader coverage of spatial information to enhance the quality of super-resolution images, these advances still require specialised optics to meet the particular excitation conditions to produce images with multiplexed PSF. As a consequence, the complexed optics and efforts in maintaining the proper alignments and system stabilities are the critical limiting factors for point-scanning super-resolution microscopes to be widely used in material science and biology labs.

Here, we report a strategy by exploring the opportunities in multi-colour emission PSF engineering in the image's Fourier domain. Using a single doughnut-shaped beam point-scanning illumination, we investigate the heterochromatic nonlinear responses and saturated fluorescence emissions of UCNPs. We develop a multi-colour Fourier domain fusion algorithm to enlarge the frequency shifting coverage of the optical system, and thereby achieve the effective super-resolution PSF by fusing Fourier components from each emission bands and processing optical transfer function (OTF) that contains the optimised spatial information.

**Results**
**PSF engineering for spatial domain subtraction**
Figure 1 presents the general concept of PSF engineering strategies in our single-doughnut-beam scanning super-resolution microscopy. It takes advantage of heterochromatic saturated emission PSFs from a multi-colour emitting probe. Instead of creating the contrast from dual-beam (de-) excitation patterns, the emission in different colour bands can have distinct power-dependent responses to a single doughnut excitation, so that to display the different emission PSF patterns, e.g. red $PSF_{Gau}$ (Gaussian PSF) and green $PSF_{Dou}$ (doughnut PSF), as shown in Fig. 1A. Due to the emission saturation effect (22), the emission doughnut PSF contains more information at high spatial frequency. This first offer the spatial domain opportunity in PSF engineering by simply subtracting the image of $PSF_{Dou}$ from the one of $PSF_{Gau}$ with an appropriate normalising coefficient, following $PSF_{FED} = PSF_{Gau} - r * PSF_{Dou}$, so that the sub-diffraction image can be obtained by a single beam scanning super-resolution microscopy (Fig. 1B & fig. S1). This approach is compatible with the typical laser scanning microscope by adding a vortex phase plate to generate a doughnut excitation beam.

Nevertheless, the process of spatial domain subtraction towards super-resolution images can introduce negative PSF components that lead to inaccurate OTF. The direct subtraction of a set of normalised data often generates areas with negative intensity values, which causes significant data loss and image distortion. Therefore, we introduce the contrast factor $r$ to adjust the strength of the subtraction. Since the subtraction method leads to the loss in critical spatial frequencies, researchers further employ a pixel assignment (24) approach to adjust the contrast factor in each pixel, but at the expense of the overall resolution.

**PSF engineering for Fourier domain fusion**
Alternatively, we can transfer the heterochromatic emission PSFs into the Fourier domain and achieve the super-resolution imaging with frequency shifting mechanism conceptualisation. As illustrated in Fig. 1C, this strategy can effectively overcome the issues of information loss and image distortion associated with the spatial domain subtraction approach, as Fourier domain OTF fusion can maximise the overall coverage of emission PSF patterns.

More specifically, by using a doughnut-shaped excitation beam, the spatial frequency components are encoded into the images by both the doughnut emission PSF and the Gaussian-like emission PSF (obtained from the over-saturated doughnut emission PSF, see the details to achieve this condition, as discussed around the results presented in Fig. 2). Benefiting from the emission saturation effect, the saturating doughnut emission PSF (Fig. 1E) contains more information at a high spatial frequency than the Gaussian-like emission PSF so that the high spatial frequency components can be captured in the extended range of the detection OTF to achieve super-resolution imaging. In the Fourier domain, the doughnut PSF has a gap in the intermediate frequency range (Fig. 1E), resulting in a deficient content loss of the intermediate spatial frequency information. The Gaussian-like PSF (Fig. 1D) can compensate the lost intermediate/low-frequency components. Therefore, heterochromatic OTF fusion of emission PSFs in Fourier domain takes advantage of both the doughnut PSF resolving power in providing the high-frequency contents and the compensation effect from Gaussian-like PSF that covers the medium frequency range (Fig. 1F).

With more details of the Fusion process described in Materials and Methods, briefly, we first perform a 2D fast Fourier transformation to convert the images that contain the doughnut PSF and the Gaussian-like PSF into Fourier domain images (fig. S2). We then segment and combine the Fourier domain images into one 'segmented Fourier image' by applying a Fourier binary mask, before the reconstruction by the inverse fast Fourier transformation (fig. S3). Figure 1G shows the line profiles of the Gaussian-like, doughnut and Fusion OTFs, illustrating the segmenting process in 1D. The amplitude indicates the resolving power at a certain spatial frequency. The cut-off frequency ($f_{cut}$) is found at the cross-point between the Gaussian-like OTF and the doughnut OTF. The segmented OTF (Fusion OTF) combines the Gaussian-like OTF (spatial frequency from 0 to $f_{cut}$) with the doughnut OTF (spatial frequency > $f_{cut}$) towards a large amplitude envelope.

To compare the image-resolving powers of the Gaussian-like emission PSF, the doughnut emission PSF, and the fused heterochromatic PSF, we perform a numerical simulation to scan different type of beams over a series of patterns consisting of diffraction limit dots of single emitters. As shown in Fig. 1H to J & fig. S4, the dots on different circles are separated with incremental distances. For the dots on the fourth ring (in green), the distance is too close for the Gaussian-like emission PSF to resolve them (Fig. 1H), while the doughnut emission PSF provides a higher resolution to solve each dot (Fig. 1I). For the dots with more sparse points on the third ring (in red), the doughnut PSF fails in providing some intermediate frequency information (see Fig. 1I), which resulted in a poorly resolved dots on the third circle. This Fourier domain fusion approach utilises all the complementary frequency contents, simultaneously acquired from the single doughnut beam scanning, which can optimise the image deconvolution process in super-resolution imaging and improve the distortion effect by taking all the frequency contents. As the final detected signal is the saturated emission with enhanced contrast to the background noise, it successfully revolves the higher frequencies information that is otherwise covered by the noise in confocal scanning mode (see Fig. 1J in the inner ring in yellow).

**Heterochromatic nonlinear optical responses from UCNPs**

We choose to use lanthanide-doped UCNPs and experimentally verify our approach. UCNPs represent an entirely new class of multi-colour emitting probes that rely on high densities of multi-photon emitters (*25*), (*26*), (*27*), (*28*), (*29*), (*30*), which up-converts near-infrared photons into visible ones. Their exceptional brightness under microscopy and nonlinear optical properties make UCNPs suitable as single-molecule probes (*31*), (*32*), (*33*), (*34*), (*35*), (*36*) and super-resolution imaging (*37*), (*38*), (*39*), (*40*), (*41*). Figure 2A displays the typical emission spectra of a 47-nm UCNP based on a $NaYF_4$ host co-doped with high-concentrations of 40% $Yb^{3+}$ sensitiser ions and 4% $Tm^{3+}$ activators ions under 980 nm excitation laser (see Materials and Methods). Benefiting from the multiple long-lived intermediate states, the sensitised photon energy can be stepwise transferred onto the scaffold energy levels of $Tm^{3+}$ emitters, which eventually facilitates the multi-photon upconversion emissions from the two-photon excited state (800 nm, $^3H_4 \rightarrow {}^3H_6$) and four-photon excited state (740 nm, $^1D_2 \rightarrow {}^3H_4$). The upconversion emission at 800 nm from the lower intermediate excited level has a lower saturation threshold compared with emissions from the higher levels, as confirmed by the nonlinear fluorescence saturation curves shown in Fig. 2B.

By taking advantage of the clear contrast in saturation intensity curves of upconversion emissions from the multi-intermediate states (*42*), we scan the sample of single UCNPs using a tightly focused doughnut illumination beam and detect from multiple emission channels, including 800 nm and 740 nm, as shown in Fig. 2C (see the experimental setup in fig. S5). We find that under the low excitation power, both channels display the doughnut-pattern PSFs. The difference in heterochromatic emission PSFs becomes significant with an increased excitation power. Because of the non-zero feature at the centre dip of doughnut beam (around 1.4%, see fig. S5), at an intense power, the 800 nm over-saturated emission PSF eventually becomes a 'Gaussian-like PSF' with two-photon upconversion emission at centre reaching the maxima (Fig. 2C). This is because the increased excitation power elevates the two-photon fluorescence at the centre to reach the maxima, while the fluorescence signals away from the centre keep at the same values since they have been saturated. In contrast, the 740 nm saturating emission PSF remains as a doughnut shape.

**Resolving single nanoparticles by Fourier heterochromatic fusion**

Next, we apply this Fourier heterochromatic fusion approach on imaging single UCNPs (Fig. 3). We first acquire the standard confocal images of 800 nm and 455 nm emission bands by scanning a standard Gaussian excitation beam (fig. S6). We then obtain the pair of contrast images at 800 nm and 740 nm by scanning a doughnut-beam (Fig. 3A & B). Neither confocal results (fig. S6) nor the over-saturated image (Fig. 3A) provide sufficient resolution to distinguish single UCNPs within the diffraction limit area. In contrast, the 740 nm saturating emission doughnut PSF carries the high-frequency information from the clusters of single nanoparticles (Fig. 3B). Using the approach of Fourier domain heterochromatic fusion of the image by the Gaussian-like (over saturated doughnut) 800 nm emission PSF (Fig. 3A, inset OTF) and the image by the saturating doughnut 740 nm emission PSF (Fig. 3B, inset OTF), as shown in Fig. 3C, we demonstrate the ability in resolving the discrete nanoparticles with a discernible distance of 120 nm. Furthermore, the magnified comparison images in Fig. 3D to G clearly illustrate the enhanced high imaging quality achieved by our Fourier domain fusion method, compared with the image deconvoluted by Gaussian-like PSF and the one processed by subtraction. As a result of the obviously different nonlinear emission responses from the multiple intermediate excited states, the varying degrees of saturating emission PSFs provide the key for the Fourier domain heterochromatic fusion of the images simultaneously obtained from the multiple emission colour bands.

As the low excitation density is essential for super-resolution nanoscopy to be used for biological applications, we further demonstrate the use of 2% $Tm^{3+}$ and 40% $Yb^{3+}$ UCNPs to reduce the excitation power requirement. According to our previous works (*18*), the relatively low doping concentration of activators will facilitate UCNPs to achieve the emission saturation at low excitation power. As shown in Fig. 3 H & I and fig. S7, under 50 mW excitation, the 800 nm emission PSF turns to a Gaussian-liked profile, while the 740 nm emission PSF remains as a doughnut shape. Using the Fourier domain heterochromatic fusion method (Fig. 3K), we can resolve the two UCNPs with a distance of 95 nm (Fig. 3L). The corresponding excitation power density of 2.75 MW/cm$^2$ is well below the photo-toxic damage threshold for the living cells (*34*), (*40*). The line profiles of the processed image of single nanoparticle, shown in Fig. 3L, further demonstrate the quantified result of the significant enhanced full width at half maximum (FWHM) of 40 nm (~$\lambda$/24) from 460 nm.

**Enhanced image quality**

Moreover, we apply our method of Fourier domain heterochromatic fusion to resolve UCNPs assembled into the various large-scale patterns. While confocal microscopy (Fig. 4A & C) cannot resolve the fine details with spacing below the diffraction limit (427 nm for 980 nm excitation), our Fourier-domain fusion method has successfully resolved the UCNPs image patterns (Fig. 4B & D). We further quantify the resolving power of Fourier domain fusion by comparing the image results of a sunflower pattern from confocal imaging (Fig. 4E), Gaussian deconvolution (Fig. 4F), doughnut deconvolution (Fig. 4G), subtraction (Fig. 4H) and Fourier domain fusion deconvolution methods (Fig. 4I). The Fourier domain fusion approach clearly presents the best image quality. In our experiment, we apply a robust decorrelation analysis (*43*) to measure the averaged image resolution directly. Figure 4J shows the decorrelation analysis for the Fourier domain fusion deconvolution images (Fig. 4I), where the original image is filtered by the high-pass Gaussian filters to emphasise the specific frequency band. These filtered images are conducted with the Pearson cross-correlation (grey curves) to find the most correlated frequencies (blue cross) between each image and its normalised low-pass image. The largest number of the most correlated frequency is the cut-off frequency ($k_c$), indicating the averaged resolution. Moreover, Fig. 4K displays the resolutions obtained from the five methods (see decorrelation analysis in fig. S8), with the Fourier domain fusion achieved the best quality in resolution (106.7 nm). Notably, the resolution by decorrelation reports the image quality, and the highest resolution is around the maximum non-zero frequency value of the image in the Fourier domain (see the magenta line). Figure 4 L to O & P to S present the detailed features of the two selected areas (Fig. 4I) by the five methods. The deconvolution of confocal cannot resolve the features below the diffraction limit. The deconvoluted doughnut image shows many artifacts, due to its inherent drawback of frequency loss. Although the subtraction compensates some artifacts, it loses frequency information, e.g. the middle point cannot be presented in Fig. 4H. According to the cross-line profiles in Fig. 4T, Fourier domain fusion shows the superior power in both resolving the fine features and managing the image artefacts. Indeed both subtraction and deconvoluted doughnut can not image out the third peak (at a distance ~ 650 nm).

We further simulate the imaging results for microtubules (*44*) to highlight the importance of compensating the missing spatial frequency regions by using the Fourier domain fusion approach (fig. S9). The high image resolutions in both transverse directions are achievable for the complexed samples, as the fine structures of microtubules can be resolved using Fourier domain fusion method, as shown in fig. S9F. We also apply the Fourier ring correlation (FRC) (*45*) method to measure the resolution from the processed data (fig. S9G), which presents the advances of our Fourier domain fusion approach in enhancing the overall image quality.

**Discussion**

To the best of our knowledge, this is the first work using the multi-colour nonlinear emission responses and processing the simultaneously obtained PSFs in the Fourier domain to enhance the spatial resolution and overall imaging quality. It maximises the capabilities of Fourier domain OTF fusion of multiple emission PSFs in the spectral regime to improve the entire imaging quality, which can recover the otherwise hidden spatial information during the single beam scanning and confocal detection process. Compared with the temporal domain modulation of excitation modality that requires switching illumination pattern (*46*) or laser mode (*47*) with dual excitations procedure, the single scan method is simple, fast and stable, and can avoid the use of the additional optical components and procedures in correcting the sample drifts between multiple sequential recordings. The single-beam scanning mode using a simplified optics setup is compatible with the standard commercial or lab-based laser scanning microscopes, and therefore may overcome the current bottleneck issue associated with the system complexity and stability. This imaging modality is compatible with point-scanning-based methods such as imaging scanning microscopy (*48*) or rescanning (*49*), which may also help to mitigate the frequency deficiency issues. To remove the additional camera, we can further develop this technique to modulate chromatic aberrations by using multi-colour phase mask (*50*). The image quality could be further improved by Fourier domain fusion of the simultaneously acquired high throughput hyper-spectrum PSFs (fig. S10). By taking advantage of these nonlinear differential responses in the hyper-spectrum domain (fig. S10D), we obtain a series of emission PSFs through the parallel multi-colour detection channels (fig. S10 B & C).

**Conclusion**

In summary, we report a single-doughnut beam scanning microscopy for super-resolution imaging. The design principle is to encode the high-frequency spatial information by PSF engineering into the multiple colour channels of the image generated from the multiple excited states of UCNPs, and decode the information by engineering the Fourier domain components of the image in each of the multi-colour channels according to the maximum coverage of OTF components. This approach, based on Fourier domain heterochromatic fusion method, opens a new perspective to perform super-resolution with minimum distortion and information loss, as it maximises the coverage of all the spatial frequency details. This strategy has a great potential in improving the resolution and artificial noise management for PSF engineering based super-resolution nanoscopy, e.g. STED microscopy (*51*), (*52*), (*53*), ground sates depletion microscopy (*8*), structure illumination microscopy (*5*) and lattice light-sheet microscopy (*54*), (*55*). Although UCNPs meet our requirement and have been successfully used in our Fourier domain heterochromatic fusion microscopy, we call for the new developments of other nonlinear optical responsive probes with multi-colour emitting properties, and the combined use of existing fluorescent probes including both molecular format of organic dyes and the inorganic nanoparticles (*56*), where preferably, the emission at each band should be highly dependent on the excitation power and their nonlinear optical response to excitation saturation.

**Materials and Methods**
**Generation of simulated data**

The details of the system and the imaging procedure for the point-scanning microscopy have been described in our recent work (*18*). Briefly, the emission is collected by an objective lens with a high numerical aperture (NA = 1.4) and focused by the tube lens onto the single-photon detector, so that the effective PSF ($h_{eff}(x, y)$) can be described as:

$$\begin{cases} h_{eff}(x, y) = h_{em}(x, y) \times h_c(x, y) \\ h_{em}(x, y) = \eta(i) \times h_{exc}(x, y) \end{cases} \quad (E.1)$$

Here $h_{em}(x,y)$ is the PSF of emission; $h_c(x,y)$ is the PSF of the confocal collection system; $h_{exc}(x,y)$ is the PSF of the excitation beam (doughnut beam); $\eta(i)$ is the excitation power dependent emission intensity curve; The full width at half-maximum (FWHM) of the intensity dip in $h_{eff}(x,y)$ represents the nanoscopy resolution. The Fourier transform of $h_{eff}$ produces an effective optical transfer function ($OTF_{eff}$).

The experimentally measured intensity distribution PSF ($h_{exp}(x,y)$) on the image plane in our system is the convolution between the $h_{eff}(x,y)$ and the spatial distribution profile ($h_{UCNP}(x,y)$) of nanoparticle as below:

$$h_{exp}(x,y) = h_{eff}(x,y) * h_{UCNP}(x,y) \qquad (E.2)$$

**FED image subtraction algorithm**

FED microscopy is an effective method that can improve the resolution of laser scanning microscopy based on the two sequential scans to obtain two images with a Gaussian PSF ($PSF_{Gau}$) and a doughnut PSF ($PSF_{Dou}$), respectively. The super-resolution image is achieved by subtracting the image obtained by the doughnut beam scan from the image obtained from the Gaussian beam scan. An appropriate normalising coefficient ($r$), typically between 0.7 and 1, is used to adjust the imaging quality. The subtraction process follows the equations:

$$PSF_{FED} = PSF_{Gau} - r * PSF_{Dou} \qquad (E.3)$$

$$F_{FED} = F_{Gau} - r * F_{Dou} \qquad (E.4)$$

where $PSF_{FED}$ is the processed effective PSF; $PSF_{Gau}$ is Gaussian PSF; $PSF_{Dou}$ is doughnut PSF; $r$ is normalising coefficient; $F_{FED}$ is the processed subtraction image; $F_{Gau}$ is the image taken with $PSF_{Gau}$; $F_{Dou}$ is the image taken with $PSF_{Dou}$.

**Fourier domain heterochromatic fusion algorithm**

Inspired by the Fourier domain method used in structured illumination microscopy (SIM), we apply the frequency shifting mechanism conceptualisation to maximise the overall coverage of emission patterns in the Fourier domain, towards the high-resolution image. Under the patterned illumination, the spatial frequency components of the sample are embedded in those of the excitation and (saturated) emission patterns, so that the high spatial frequency information of the sample is moved into the larger range of the detection OTF.

Following a similar derivation reported (46), we develop a Fourier domain heterochromatic fusion method to alleviate the frequency deficiency at certain frequency points as well as the sectional deficiency induced by the saturation effect. The procedure is given by

$$OTF_{eff} = OTF_{Gau} \times mask_{Gau} + OTF_{Dou} \times mask_{Dou} \times r_1 \qquad (E.5)$$

$$F_{eff} = F_{Gau} \times mask_{Gau} + F_{Dou} \times mask_{Dou} \times r_1 \qquad (E.6)$$

where $OTF_{Gau}$ and $OTF_{Dou}$ are the Fourier transform of the corresponding $PSF_{Gau}$ and $PSF_{Dou}$, respectively. The $mask_{Gau}$ and $mask_{Dou}$ are low/high (the white part is pass) Fourier frequency filters, respectively. $r_1$ is a variable that modifies the ratio of these fused components for achieving the optimal synthetic system $OTF_{eff}$. $F_{Gau}$ and $F_{Dou}$ are the acquired images with the corresponding $PSF_{Gau}$ and $PSF_{Dou}$, respectively.

**Synthesis of $Tm^{3+}$-$NaYF_4$ nanoparticles**

The $Tm^{3+}$-$NaYF_4$ UCNPs were synthesised using a co-precipitation method (37). In a typical procedure, a methanol solution (1 mmol) of $YCl_3 \cdot 6H_2O$, $YbCl_3 \cdot 6H_2O$ and $TmCl_3 \cdot 6H_2O$ together with oleic acid (6 mL) and 1-octadecene (15 mL) was added to a 50 ml three-neck round-bottom flask under vigorous stirring. The resulting mixture was heated at 150 °C for 30 mins to form lanthanide oleate complexes. The solution was cooled down to room temperature. Subsequently, a methanol solution (6 mL) containing NaOH (2.5 mmol, 100 mg) and $NH_4F$ (4 mmol, 148 mg) was added and stirred at 50 °C for 30 mins, and then the mixture was slowly heated to 150 °C and kept for 20 mins under argon flow to remove methanol and residual water. Next, the solution was quickly heated at 300 °C under an argon flow for 1.5 h before cooling down to room temperature. The resulting core nanoparticles were precipitated by addition of ethanol, collected by centrifugation at 9000 rpm for 5 min, the final $NaYF_4$: $Yb^{3+}$, $Tm^{3+}$ nanocrystals were redispersed in cyclohexane with 5 mg/mL concentration after washing with cyclohexane/ethanol/methanol several times.

**Measurements and Characterisations**

The morphology of the nanoparticles was characterised via transmission electron microscopy (TEM) imaging (Philips CM10 TEM with Olympus Sis Megaview G2 Digital Camera) with an operating voltage of 100 kV. The samples were prepared by placing a drop of a dilute suspension of nanocrystals onto copper grids.

**Sample preparation**

For preparing a sample slide, a coverslip was washed with pure ethanol by ultrasonication and air-dried. 5 µl UCNPs suspension (diluted to 0.01 mg/ml in cyclohexane) was dropped onto the coverslip. After being air-dried, the coverslip was covered by a clean glass slide.

**Optical setup**

All the measurements have been performed on a home-built microscopy system equipped with a 3-axis closed-loop piezo stage (stage body MAX311D/M, piezo controller BPC303; Thorlabs) and a vortex phase plate (VPP, VPP-1a, RPC Photonics). UCNPs are excited by a polarisation-maintaining single-mode fibre-coupled 980 nm diode laser (BL976-PAG900, controller CLD1015, Thorlabs). The first half-wave plate (HWP, WPH05M-980, Thorlabs) and a polarised beam splitter (PBS, CCM1-PBS252/M, Thorlabs) are employed to precisely adjust the excitation power by rotating HWP electronically. The purpose of the second HWP was to turn the polarisation from horizontal (P polarised) to vertical (S polarised). A doughnut-shaped point spread function (PSF) at the focal plane is generated by a VPP. Confocal scanning is acquired without the VPP via the auxiliary two flexible mirrors. After collimation, the excitation beam is reflected by the short-pass dichroic mirror (DM1, T875spxrxt-UF1, Chroma), and focused through a high numerical aperture objective (UPlanSApo, 100×/1.40 oil, Olympus) to the sample slide. A quarter-wave plate (QWP, WPQ05M-980, Thorlabs) is adopted to transform the excitation beam from linear polarisation to circular polarisation to obtain optical super-resolution images. Photoluminescence is collected by the same objective and split from the excitation beams by DM1. The emission signals are separated by the short-pass dichroic mirror (DM2, T785sp, Chroma), and purified by the band-pass filter (BPF1, ET740/60M, Chroma) and the band-pass filter (BPF2, ET805/20M, Chroma), respectively. The emission can be coupled into multimode fibre (MMF1 & MMF2, M42L02, Thorlabs), then detected by a single-photon counting avalanche photodiode (SPAD1 & SPAD2, SPCM-AQRH-14-FC,

Excelitas), or by a miniature monochromator (iHR550, Horiba) for measuring upconversion emission spectra. Typical excitation powers for the recording of super-resolution images varied from 5 mW to 150 mW. All powers are measured at the back aperture of the objective lens. Pixel dwell times are adjusted as 1 ms.

**The pattern on the etched silicon wafer**

A standard Electron Beam Lithography (EBL) technique was used to fabricate patterns. The wafer was firstly cleaned by using chemicals through the following three steps: cleaning using acetone and isopropanol, and drying with nitrogen gas, respectively. After this, we spin-coated a layer of poly (methyl methacrylate) (PMMA) EBL resist onto the wafer. Then, we make use of a Raith 150TWO to write the pattern on the wafer. Following PMMA development, the nano-wells of 150 nm in diameter and 100 nm in depth were then etched using Oxford Instruments Plasmalab System 100 reactive ion etching system. After the remaining EBL resist was removed using acetone, the cyclohexane suspension of UCNPs was cast onto the etched silicon chip and dried under ambient condition. The nano-wells were spontaneously filled with the UCNPs.

**Acknowledgments**

The authors thank Joanna Szymanska and Andrew See for the pattern fabrication, performed at the NSW Node of the Australian National Fabrication Facility. **Funding:**. The authors acknowledge the financial supports from the UTS Chancellor's Postdoctoral Research Fellowship (PRO18-6128), the Australian Research Council (ARC) DECRA fellowship (DE200100074 - F.W.), the ARC Discovery Project (DP190101058 - F.W), National Natural Science Foundation of China (NSFC, 61729501), Major International (Regional) Joint Research Project of NSFC (51720105015), Science and Technology Innovation Commission of Shenzhen (KQTD20170810110913065), and Australia China Science and Research Fund Joint Research Centre for Point-of-Care Testing (ACSRF658277, SQ2017YFGH001190). C.C., B.L., Y.L., J.L. and X.S. acknowledge the financial support from China Scholarship Council (C.C.: No. 201607950009; B.L.: No. 201706020170; Y.L.: 201607950010; J.L.: No. 201508530231; X.S.: 201708200004). **Author contributions:** C.C., F.W. and D.J. conceived the project and designed experiments. C.C., Y.L., F.W. and X.S. conducted the optical setup and performed the optical experiments. B.L. wrote the Fourier fusion algorithm. C.C. and F.W. built the optical simulation model. Y.L. and F.W.

built the rate-equation model. J.L. synthesised the upconversion nanoparticles. C.C., F.W. B.L. and D.J. analysed the results, prepared the figures and wrote the manuscript. All authors participated in the editing of the manuscript. **Competing interests:** The authors declare no competing interests. **Data and materials availability:** All data needed to evaluate the conclusions in the paper are present in the paper and/or the Supplementary Materials. Additional data related to this paper may be requested from the authors.

**Figures**

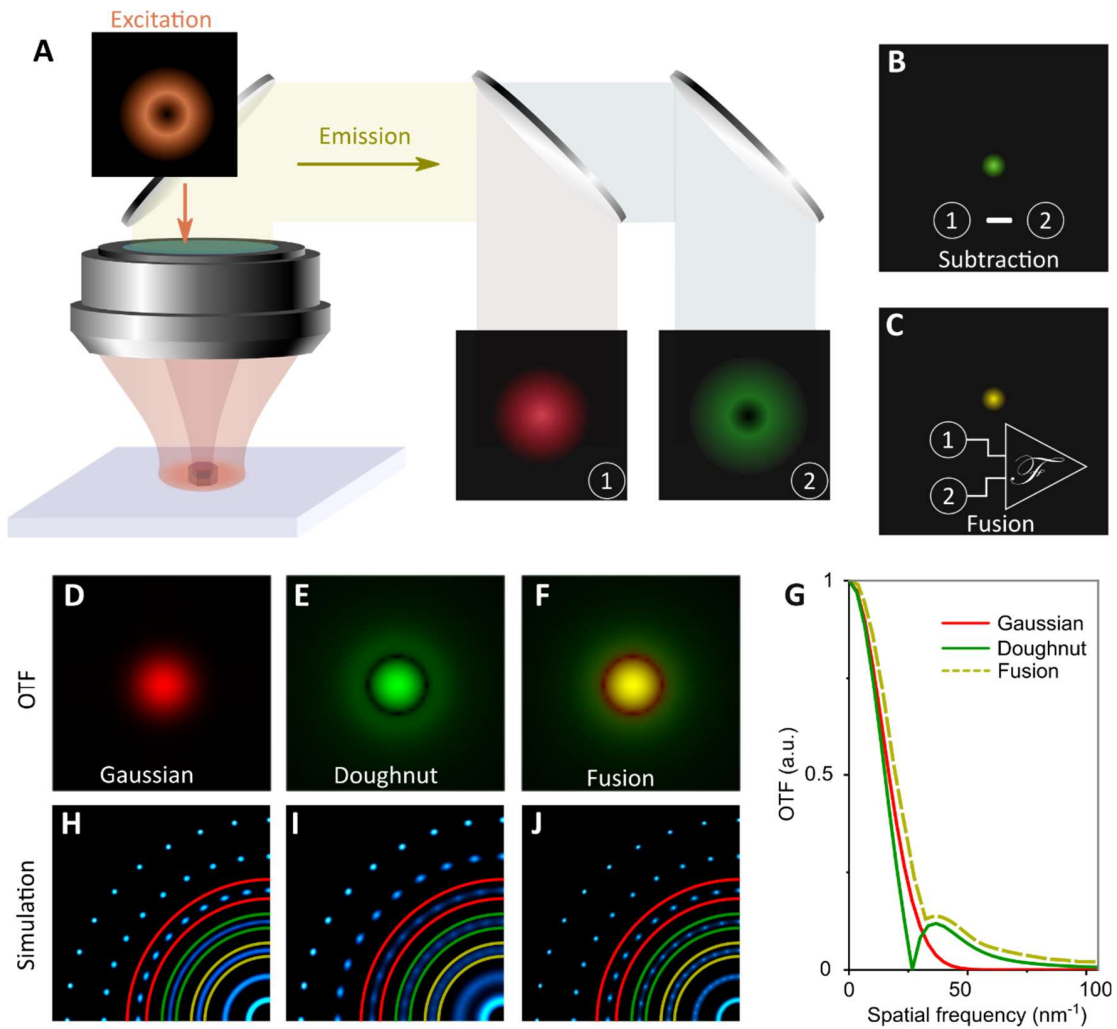

**Fig. 1. Concepts of heterochromatic PSF engineering.** (**A**) One doughnut illumination beam generates two power-dependent emission PSF patterns of Gaussian in red (1) and doughnut in green (2). (**B**) PSF engineering by subtracting the doughnut PSF from Gaussian PSF in the spatial domain. (**C**) PSF engineering by fusing the two PSFs in the Fourier domain. (**D** to **F**) The OTFs of Gaussian PSF, doughnut PSF, and OTF fusion result of Gaussian and doughnut PSFs. The zero spatial frequency starts from the centre of each image, and higher orders of frequencies increase radially. (**G**) Normalised centre cross-section profile of OTFs corresponding to (F). (**H** to **J**) Simulated images of a series of dots of PSFs separated by different distances and super-resolved by Gaussian PSF (D), doughnut PSF (E), and synthesised PSFs (F). Iterations in (H to J) are all 180.

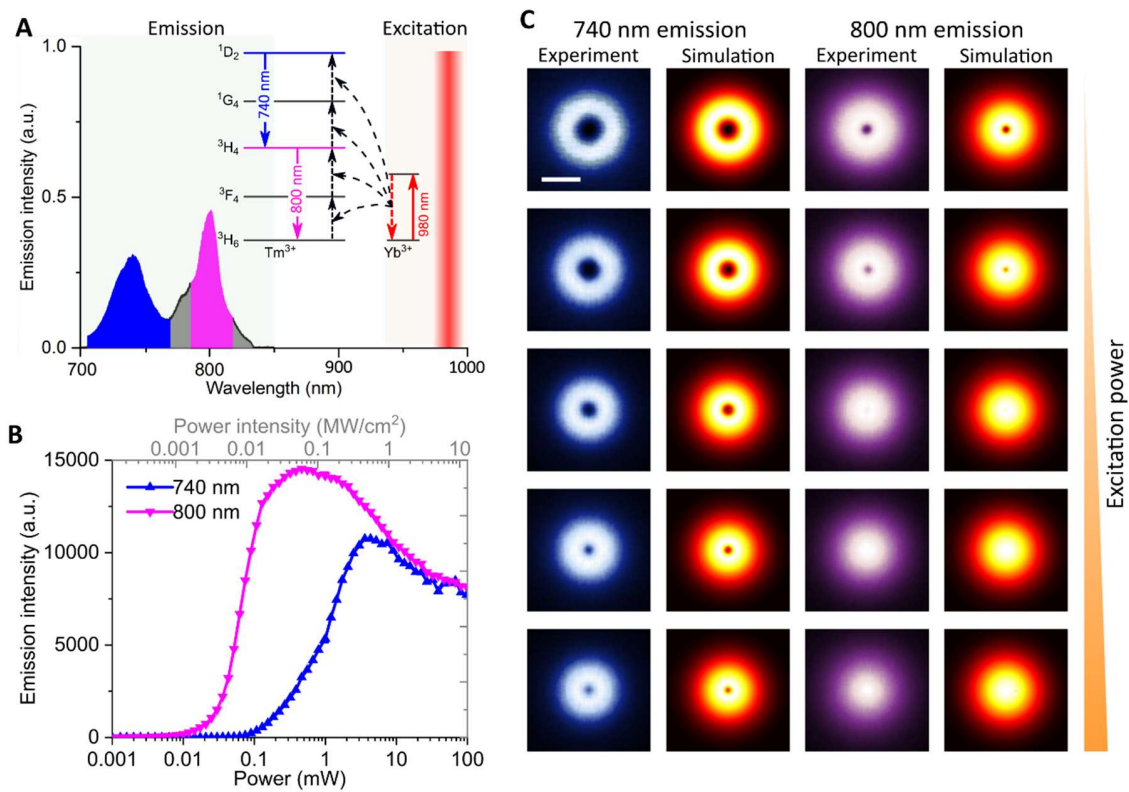

**Fig. 2. Heterochromatic emission saturation contrast in UCNPs**. (**A**) The upconversion emission spectrum of a typical 47 nm single nanoparticle (NaYF$_4$: 40% Yb$^{3+}$, 4% Tm$^{3+}$) upon 980 nm excitation laser at a power of 100 mW. Inset is a simplified energy level and upconversion process of Yb$^{3+}$ and Tm$^{3+}$ co-doped UCNPs. The multi-photon near-infrared (NIR) upconversion emissions mainly from the two-photon excited state (800 nm, $^3H_4 \rightarrow {}^3H_6$) and four-photon excited state (740 nm, $^1D_2 \rightarrow {}^3H_4$). (**B**) The two distinct power-dependent saturation intensity curves of the 800 nm and 740 nm emissions. (**C**) The experimental and simulation results of the power dependent PSF patterns of two emission bands from a single UCNP with excitation powers of 5 mW, 20 mW, 50 mW, 100 mW, and 150 mW. Pixel size, 10 nm. Scale bar is 500 nm.

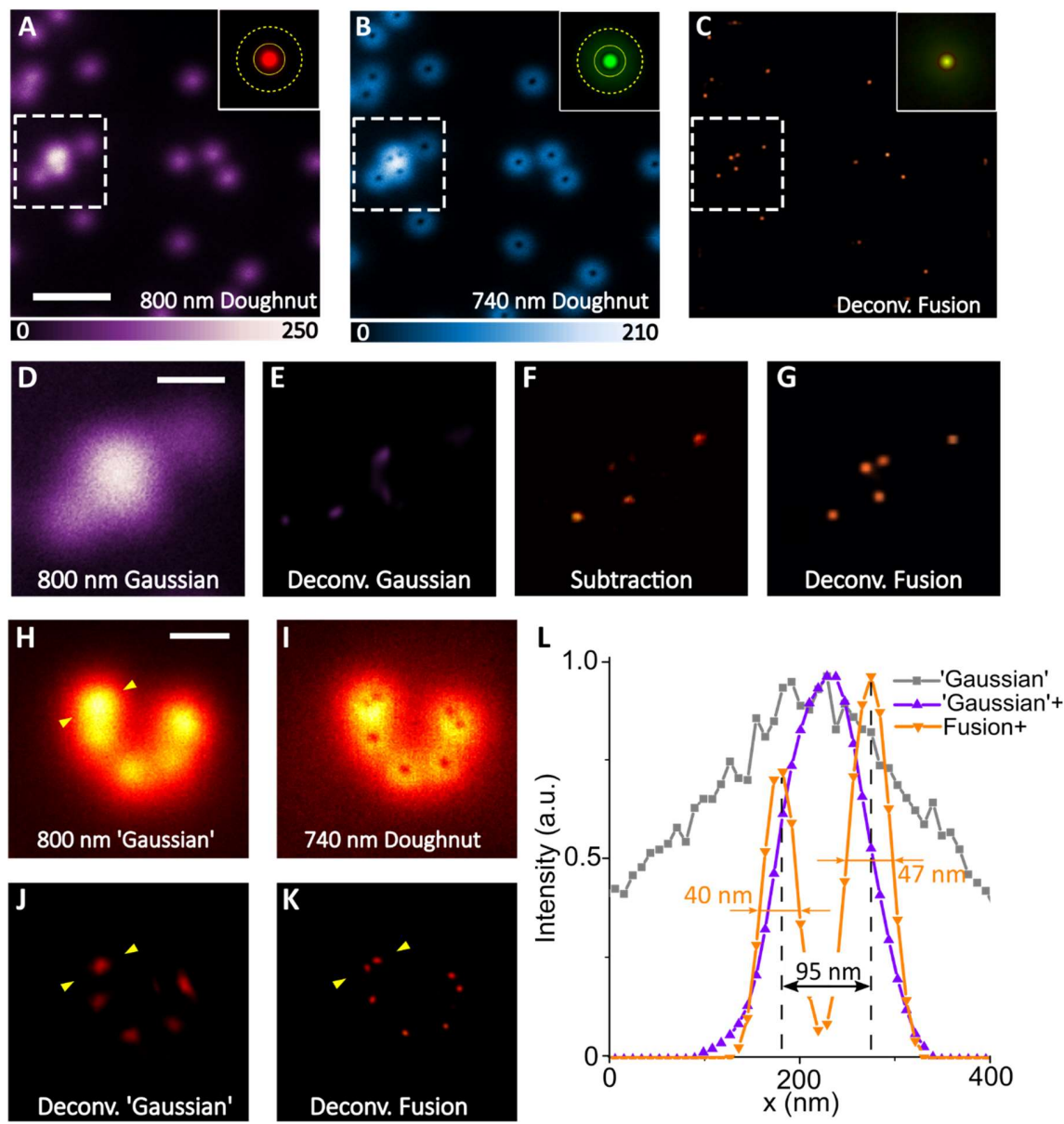

**Fig. 3. Resolving the signal UCNPs in sub-diffraction volume.** (**A**) The 800 nm emission band image of UCNPs under a 980 nm doughnut beam excitation (150 mW). With UCNPs' 800 nm emission being over-saturated, the emission PSF shows a Gaussian-like profile. Inset is the corresponding OTF. (**B**) The 740 nm emission band image of UCNPs under the same 980 nm doughnut beam excitation, showing the doughnut emission PSF. Inset shows the corresponding OTF. (**C**) The super-resolution imaging result by Fourier domain fusing the OTFs of (A) with (B). Inset is the fused OTF. (**D** to **G**) The magnified area of interest to illustrate the comparison imaging results using the various image process algorithms, including Richardson-Lucy deconvolution with (E) Gaussian PSF (Deconv. Gaussian), (F) subtraction of the doughnut image from the Gaussian image (Subtraction), and (G) the Fourier domain fusion (Deconv. Fusion), respectively. (**H** & **I**) The 800 nm and 740 nm emission band image under 50 mW doughnut beam, respectively. (**J** & **K**) The corresponding deconvolution images in (H & I), respectively. (**L**) Line profiles of two nearby UCNPs from (H to K). Pixel dwell time, 1 ms. Pixel size, 10 nm. The scale bars are 1.5 μm in (A & C), and 500 nm in (D to K).

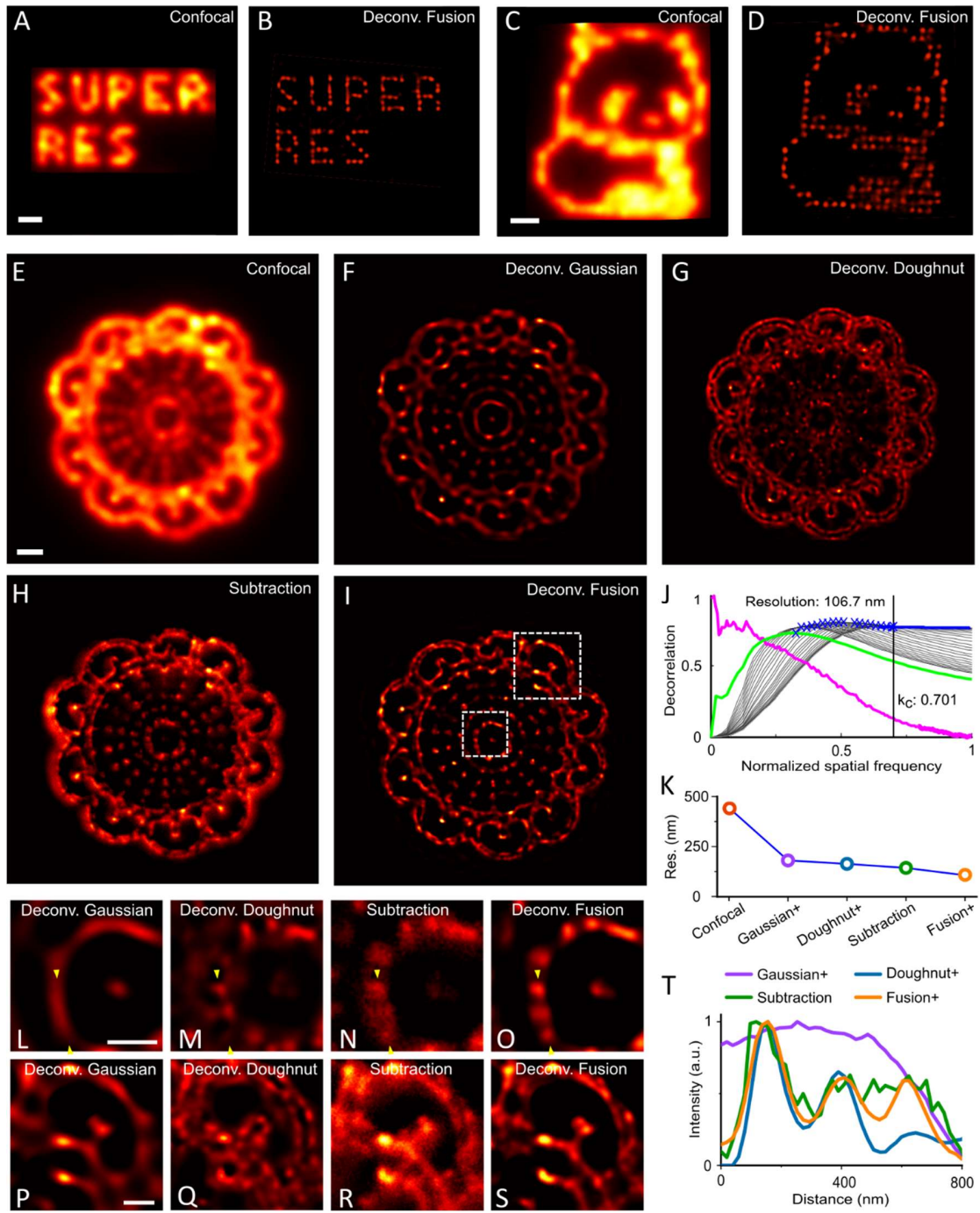

**Fig. 4. The overall enhanced image quality in super-resolution.** (**A** and **B**) The confocal and the Fourier domain heterochromatic fusion deconvoluted images of the 'super res' text pattern. (**C** and **D**) The confocal and the Fourier domain heterochromatic fusion deconvoluted images of a panda pattern. (**E** to **I**) The image results of a sunflower pattern from five processing methods: (E) Confocal imaging; (F) Gaussian deconvolution; (G) Doughnut deconvolution; (H) Subtraction; (I) Fourier domain heterochromatic fusion deconvolution. (**J**) The decorrelation analysis for the image in (I). The resolution is 2*(pixel size)/$k_c$, where $k_c$ is expressed in normalised frequencies. (**K**) The resolution comparison for the five methods in (E to I). (**L** to **O** & **P** to **S**) The magnified regions in (I). (**T**) Corresponding cross-line profiles in (L to O). The patterns were generated by using the electron beam lithography for etching 150 nm holes on a silicon substrate. The iteration number for all the deconvolution process is 70. Pixel dwell time, 1 ms. The scale bars are 1 μm in (A to I), and 500 nm in (L to S).

Supplementary Materials for

# Heterochromatic nonlinear optical responses in upconversion nanoparticles for point spread function engineering

Chaohao Chen[1], Baolei Liu[1], Yongtao Liu[1], Jiayan Liao[1], Xuchen Shan[1], Fan Wang[1,2]*, and Dayong Jin[1,3]*

[1]Institute for Biomedical Materials & Devices (IBMD), Faculty of Science, University of Technology Sydney, NSW 2007, Australia

[2]School of Electrical and Data Engineering, Faculty of Engineering and Information Technology, University of Technology Sydney, Ultimo 2007, Australia

[3]UTS-SUStech Joint Research Centre for Biomedical Materials & Devices, Department of Biomedical Engineering, Southern University of Science and Technology, Shenzhen, China

*Correspondence to fan.wang@uts.edu.au, dayong.jin@uts.edu.au

**The Supplementary Materials include:**

    Supplementary Text
    Fig. S1 to S10

# Supplementary text

The concept of this work is based on extracting the distinct features in nonlinear photon responses of multi-colour emission probes and engineering the point spread function (PSF) of the single probe in both the spatial and Fourier domains. In this section, we describe the working algorithms underpinning the methods of Fluorescence emission difference (FED) image subtraction and Fourier domain heterochromatic fusion, respectively. We evaluate their performances in nanoscopy through theoretical simulations.

## FED image subtraction algorithm

In fig. S1, rather than obtaining the two images from two scans, we develop a super-resolution approach based on the two emission PSFs (Gaussian-like and doughnut) by exciting the nonlinear multi-colour probe. The simulation results are shown in fig. S1B to F. Compared with the image obtained from $PSF_{Gau}$ (fig. S1 B & E), the resolution of the processed image by subtracting the emission doughnut PSF from the Gaussian-like emission PSF has been clearly enhanced (fig. S1 E & F).

Although our new method has simplified two scans based modality of FED, the main shortcomings of both subtraction-based modalities suffer from the image distortion of artefacts and critical information loss. The subtraction balances the intensities of the initial images, and the direct subtraction of normalised data sets often generates areas with negative intensity values, which is a potential source of data loss. When an appropriate normalising coefficient (*r*) is introduced to adjust the strength of the subtraction, the *r* as a constant for the entire imaging area will cause over-subtracting and the resultant data loss. As the processed image is shown in fig. S1F, the subtraction method creates an absence in critical spatial frequencies. Moreover, the processed image shows the dots with different sizes, apparently as the artefacts.

## Fourier domain heterochromatic fusion algorithm

The details in the image process are shown in fig. S2 & S3. Firstly, we obtain the effective PSF for the fusion process by using the Gaussian-like and doughnut PSF (fig. S2). In the beginning, the emission spectrum was collected to determinate the multiple emission bands (emission wavelength) of the fluorescent probes. As each emission band has different emission colour, the two power-dependent curves were obtained by using wavelength band-pass filters or a dichroic mirror. The two different emission PSFs (Gaussian-like and doughnut) were measured by employing an intense doughnut-shaped beam to scan a single fluorescent probe. Both of the raw images may contain complementary information, which can be analysed by the OTF spectrum from each emission band through the fast Fourier transformation (FFT). By setting the cut-off frequency, the two different spectrums are fused into a single compound spectrum to yield an $OTF_{eff}$, referred to as the "segmented Fourier spectrum". For the PSF reconstruction, the inverse fast Fourier transformation (iFFT) was applied to yield the images with the emission $PSF_{eff}$. Meanwhile, the low/high-frequency pass masks were obtained for the next imaging processing step.

With the effective PSF and frequency masks, we can process the images by the Fourier domain heterochromatic fusion method (fig. S3). Firstly, the two images of different colour emission were obtained by using a focused doughnut-shaped beam to scan the specimen. Then the two images were converted to the Fourier domain through FFT. With the frequency filter masks (masks are from fig. S3), the information in the Fourier domain was selected by applying the low-pass (LP) mask for the image of Gaussian-like and high-pass (HP) mask for the counterpart. Subsequently, a new optimal Fourier domain was achieved by fusing these two selected Fourier domains. The fused image can be obtained by using iFFT. Finally, the high-quality image is achievable for the fused image by Richardson-Lucy deconvolution with the $PSF_{eff}$ ($PSF_{eff}$ is from fig. S3).

Figure S4M compares the difference in frequency components between OTFs obtained by two emission PSFs. Obviously, $PSF_{Dou}$ generates more contents at a high spatial frequency than that generated by $PSF_{Gau}$. Larger content of high frequencies in OTF helps to recover high spatial frequency components in the image through the deconvolution algorithm.

However, the emission $PSF_{Dou}$ has a gap in the mediate frequency range, resulting in a deficiency at the mediate to low spatial frequencies. While this loss can be compensated by the OTF components generated from $PSF_{Gau}$ (fig. S4M). The process to merge different OTF components is called Fourier domain fusion. The multiple OTF components from different PSF components can be achieved by either multiple scans with different PSFs, as in FED method, or by the multi-colour emission PSFs from fluorescence probes with power-dependent nonlinear responses to the excitation doughnut PSF as presented in this work.

To evaluate the imaging performance using PSF engineering, we conduct a numerical experiment by designing a series of patterns consisting of emitters with varying distances, as shown in fig. S4. For the circles assembled with closer emitter-emitter distance, the image acquired with $PSF_{Dou}$ can clearly resolve the dots, indicating a higher achievable resolution compared with that achieved in the $PSF_{Gau}$ mode. For the ones with more sparse points, the images obtained in the $PSF_{Dou}$ scanning mode suffers from the distortion effect, as it lacks some mediate frequencies. The maximum spatial information in focus plane can be obtained by merging the maximum Fourier components from different emission PSFs, so that we design a Fourier binary mask to select and fuse these components (fig. S4 H & I), to yield the overall improved imaging quality. As the fusion technique utilises all the spatial frequency information revolved by the two PSFs, complimentary high-resolution details can be collected without distortion together contribute to the high-quality images, showing a clear contrast to the confocal images. This approach also avoids the use of extra optical elements, as it simultaneously yields different emission PSFs from different colours. The complementary frequency contents further facilitate the deconvolution process in super-resolution imaging.

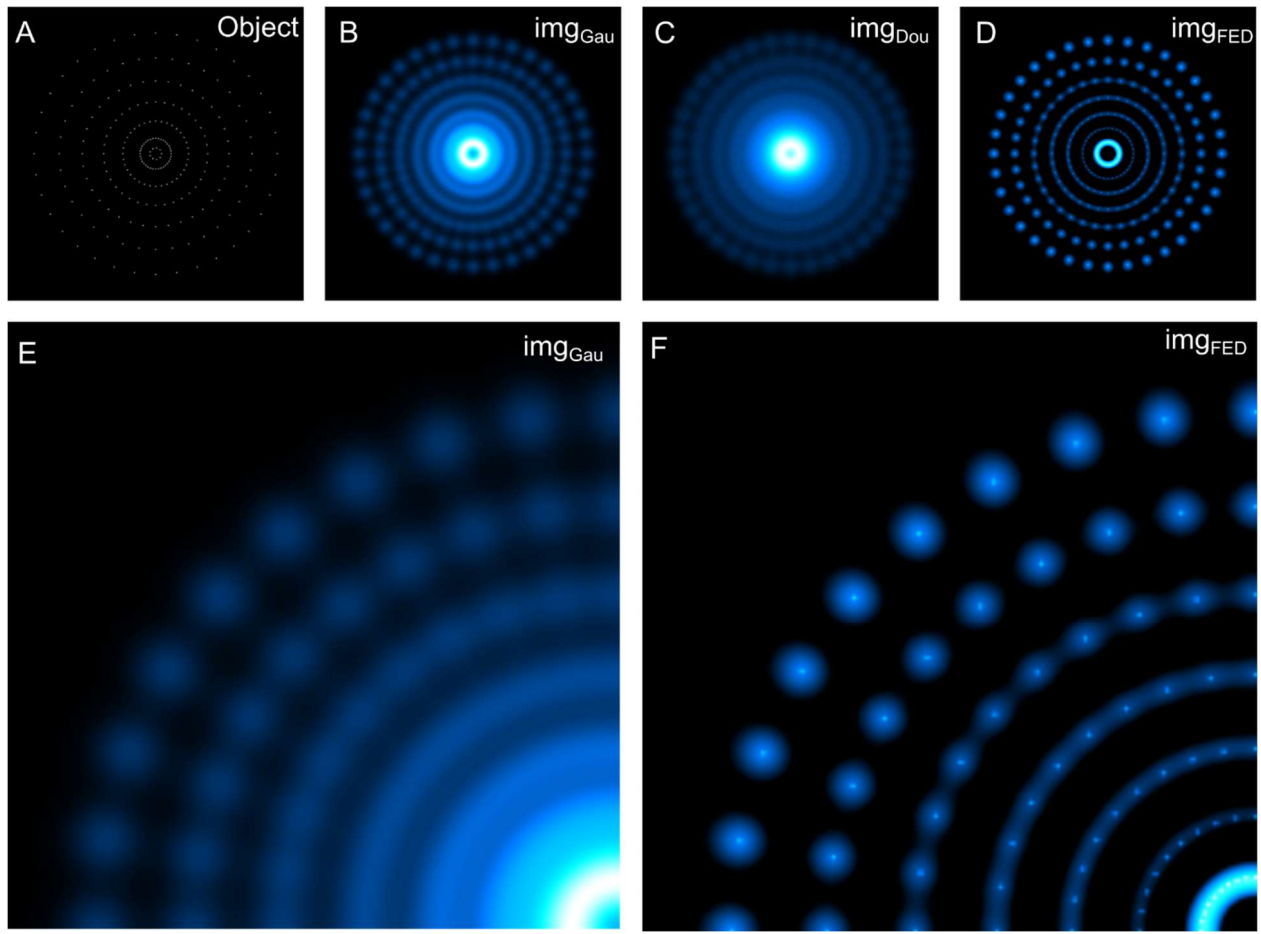

**Fig. S1. Simulation result by FED image subtraction.** (A) The imaging object: a series of patterns of diffraction limit dots of single emitters. (B) The scanning image by Gaussian PSF and (C) Doughnut PSF. (D) The processed result by subtracting the image in (C) from the in (B). (E) The magnified image in (B). (F) The magnified image in (D). In this simulation, the appropriate normalizing coefficient *r* is 0.9.

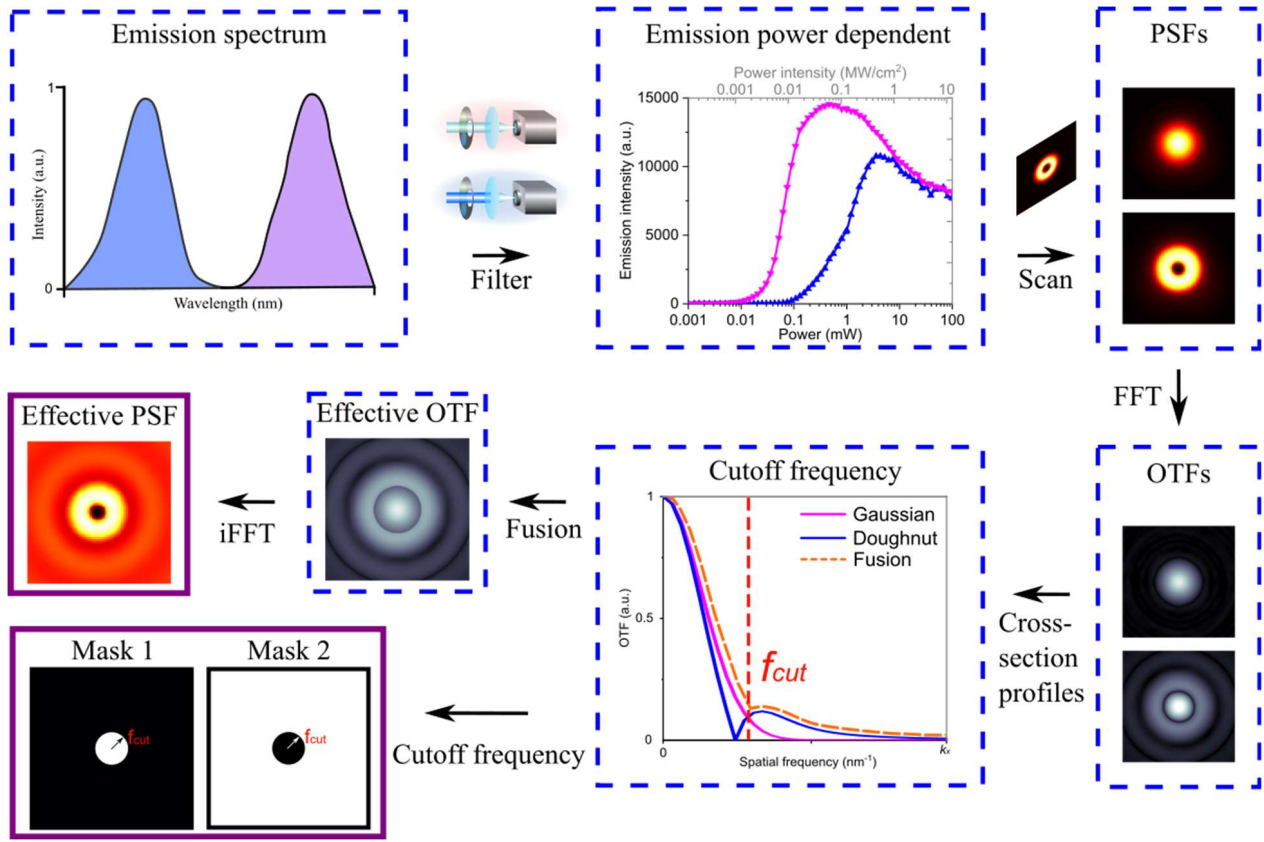

**Fig. S2. Effective PSF for the fusion process by using the Gaussian and doughnut PSF.** By selecting the cut-off frequency, we can obtain the $PSF_{eff}$ and the low/high-frequency pass masks for imaging processing.

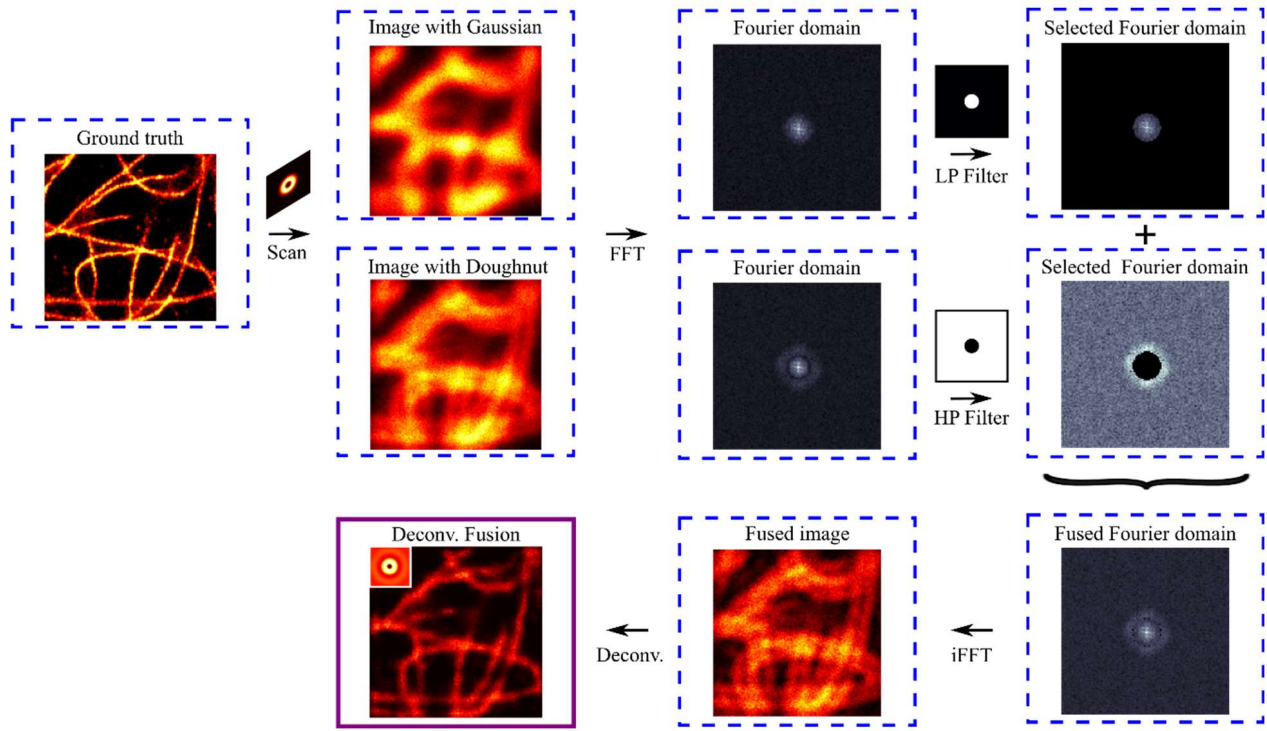

**Fig. S3. The flow chart of the Fourier domain heterochromatic fusion method.** The high-quality image is achievable for the fused image by Richardson-Lucy deconvolution with the $PSF_{eff}$.

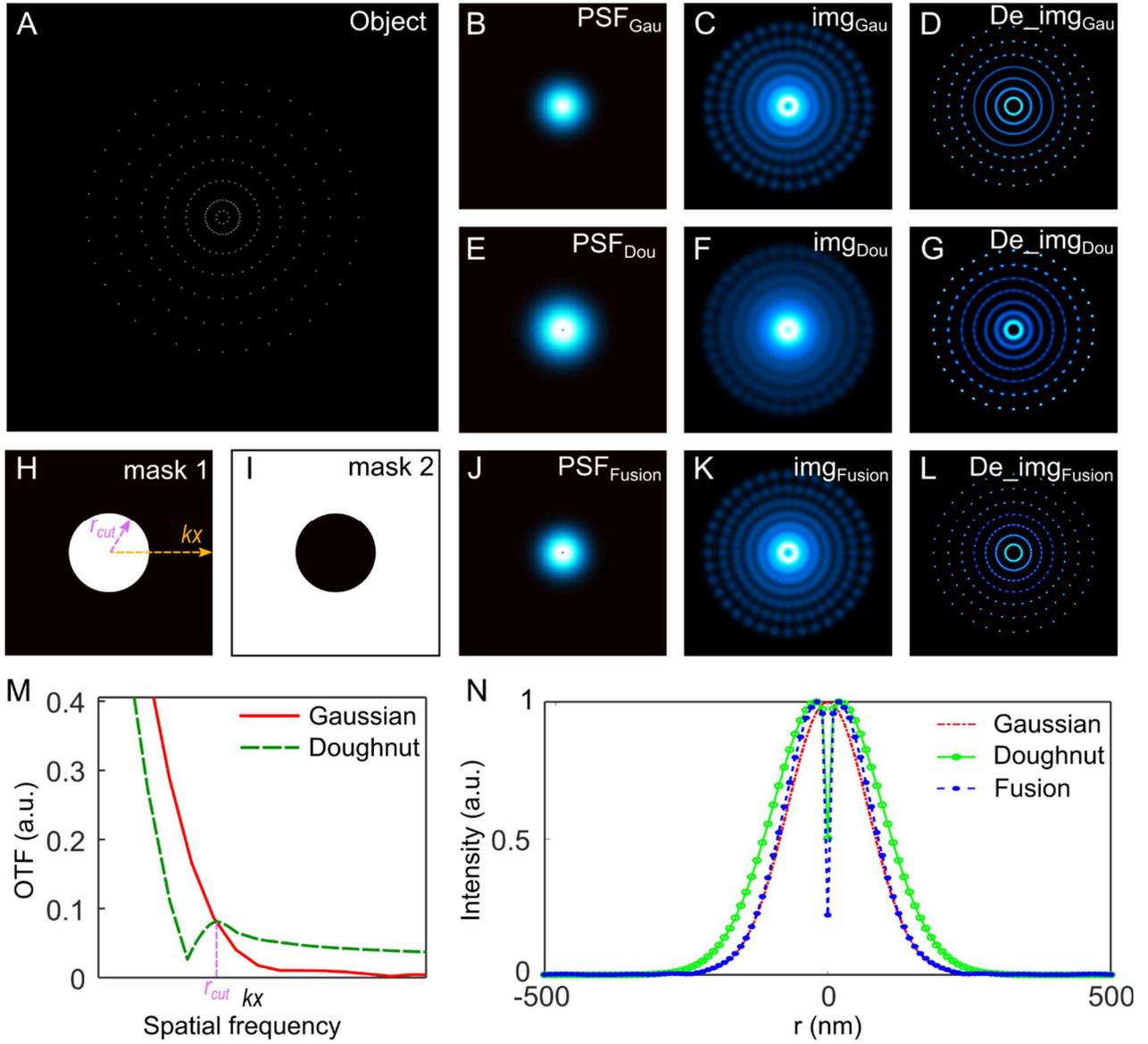

**Fig. S4. A numerical study using the Fourier domain heterochromatic fusion algorithm to overcome the issues associated with frequency deficiency and imaging distortion.** (**A**) The original object with a series of patterns of diffraction limit dots of single emitters. (**B**) Gaussian PSF. (**C**) The image obtained by using Gaussian PSF. (**D**) Richardson-Lucy (RL) deconvolution result of the image in (C) with the Gaussian PSF in (B). (**E**) Doughnut PSF. (**F**) The image obtained using Doughnut PSF. (**G**) RL deconvolution result of the image in (E) with the Doughnut PSF in (F). (**H**) Mask1 corresponds to a circular low-pass filter used to block the high-frequency component above $r_{cut}$. The orange arrow of $k_x$ denotes the direction of the spatial-frequency increase. (**I**) Mask 2 is a high-pass filter that permits the high-frequency component above $r_{cut}$. (**J**) The corresponding fusion PSF after implementing the proposed algorithm. (**K**) The image obtained using the fusion algorithm. (**L**) RL deconvolution result by deconvolving the image in (K) with the fusion PSF in (J). (**M**) OTF profiles of Gaussian (red solid) and doughnut (green dash), respectively. $r_{cut}$ is a point selected at the intersection of Gaussian and Doughnut profiles. (**N**) The corresponding PSF profiles of Gaussian (red), Doughnut (green) and Fusion (blue), respectively. The iterations of the simulation are 180.

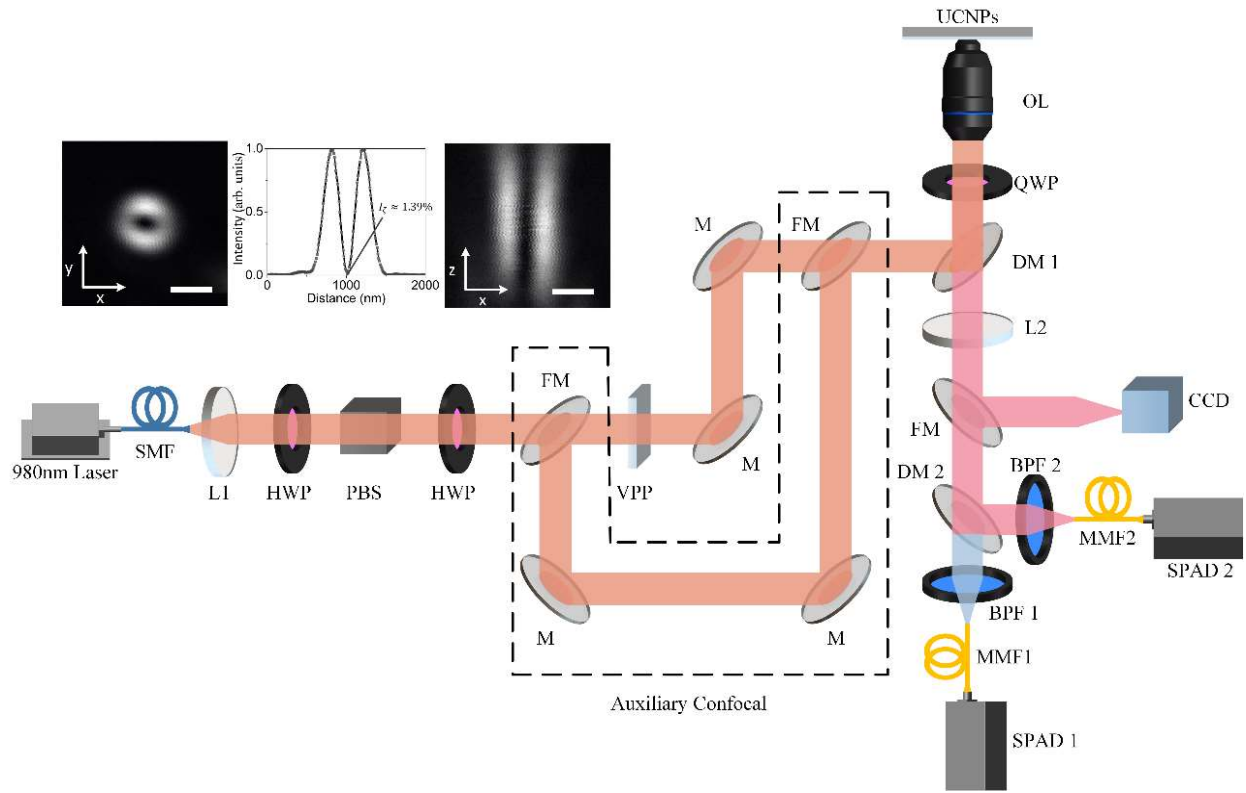

**Fig. S5. Schematic setup of the single beam scanning super-resolution nanoscopy** (SMF, single-mode fibre; MMF1 and MMF2, multi-mode fibre; L1, collimation lens; L2, collection lens; HWP, half-wave plate; QWP, quarter-wave plate; PBS, polarised beam splitter; VPP, vortex phase plate; M, mirror; FM, flexible mirror; DM1 and DM2, dichroic mirror; OL, objective lens; BPF1 and BFP2, bandpass filter; SPAD1and SPAD2, single-photon avalanche diode; CCD, charge-coupled device). The dotted portion is designed for auxiliary confocal with two flexible mirrors to bypass the VPP in the main optical path. Inset shows the PSF of the excitation beam measured by the scattering signal from a 100 nm gold nanoparticle in reflection (path not shown). The $I_\zeta$ (ratio value of the intensity at the doughnut centre to the max intensity of the beam) is measured as 1.39%. Scale bars: 500nm.

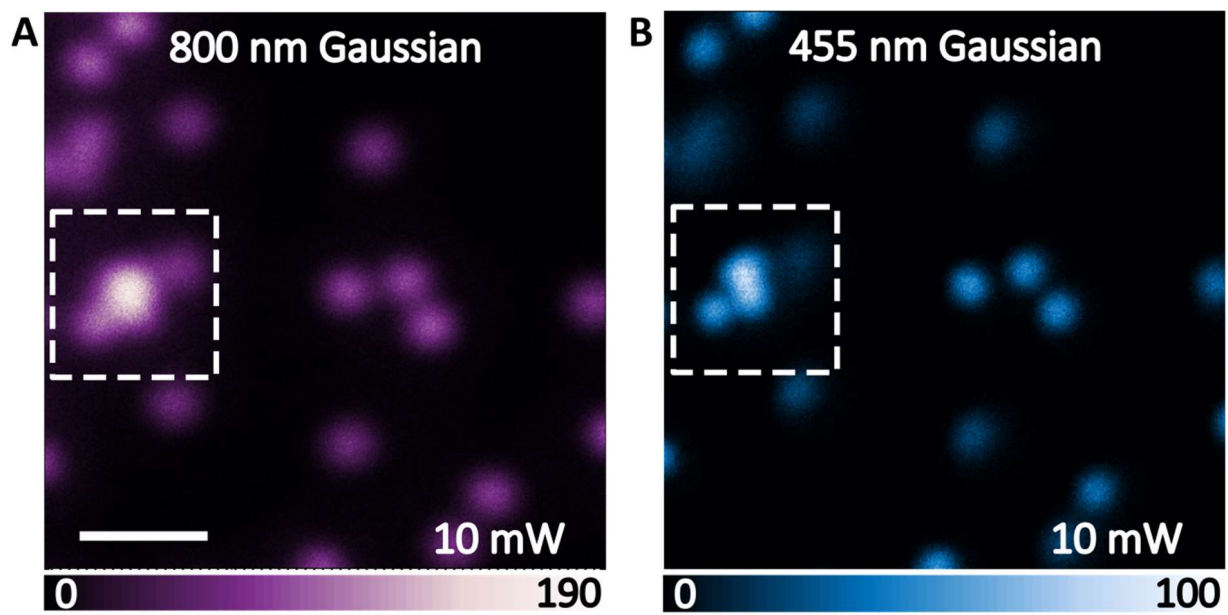

**Fig. S6. Confocal imaging of the UCNPs under 980 nm Gaussian beam excitation.** (**A**) 800 nm emission band image. (**B**) 455 nm emission band image. Pixel dwell time, 1 ms. Pixel size, 10 nm. The scale bars are 1.5 μm.

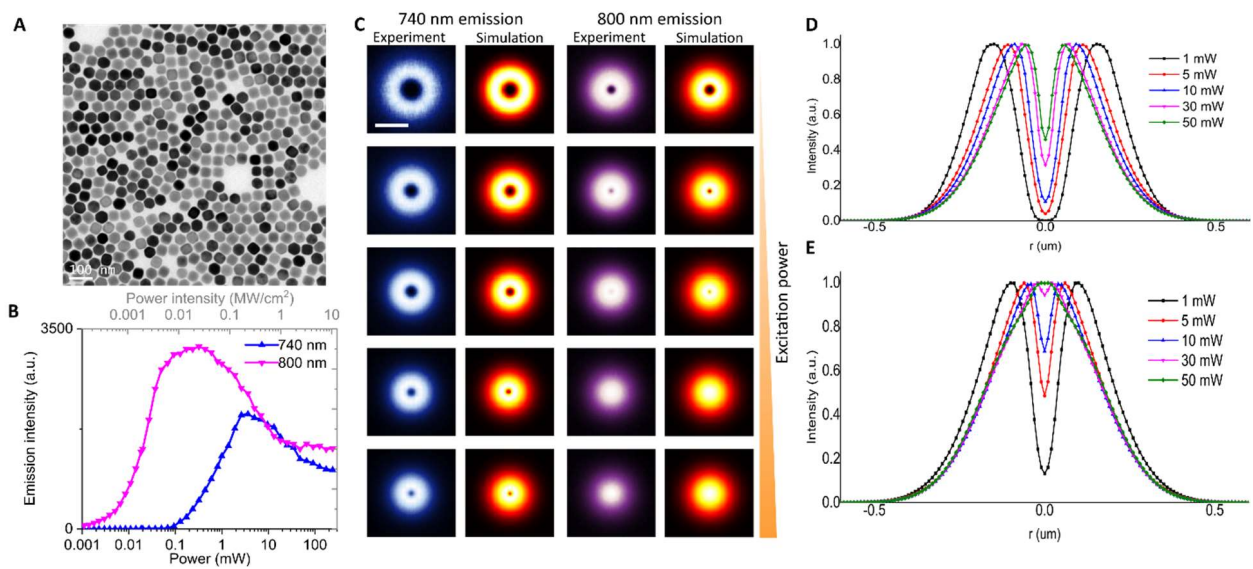

**Fig. S7. Heterochromatic emission saturation contrast for the low lanthanide-doped UCNPs**. (**A**) TEM images of the nanoparticles NaYF$_4$: 40% Yb$^{3+}$, 2% Tm$^{3+}$. The average size is around 50 nm. (**B**) The emission power-dependent curves. (**C**) The experimental and simulation results of the power dependent PSF patterns of two emission bands from a single UCNP when increasing the laser power from 1 mW to 50 mW. (**D** & **E**) The corresponding cross-section profiles of simulated 740 nm emission PSFs and 800 nm emission PSFs in (C), respectively. Scale bar is 100 nm in (A), and 500 nm in (C).

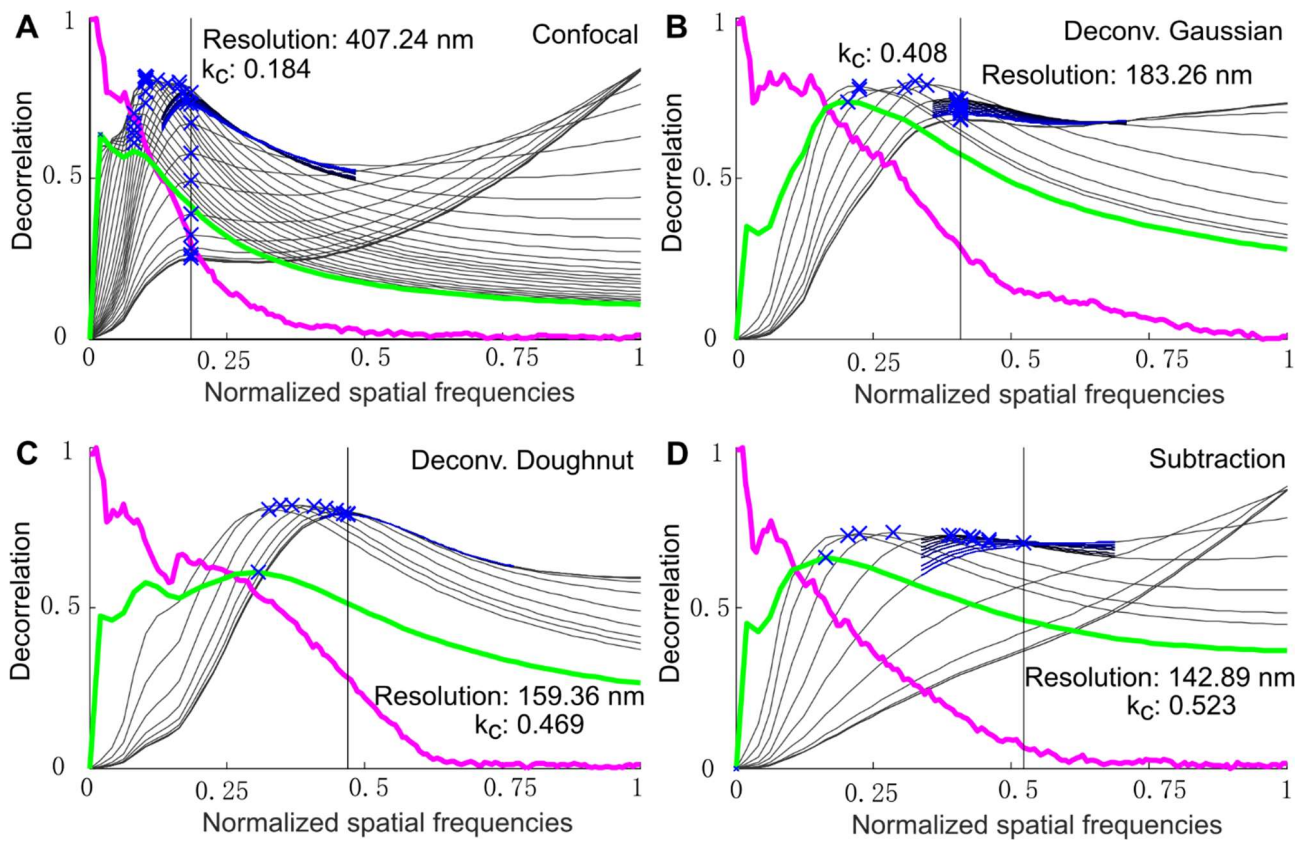

**Fig. S8. The decorrelation analysis for the image processed by different methods.** (**A**) The result of the confocal image. (**B**) The result of the image by Gaussian deconvolution. (**C**) The result of the image by doughnut deconvolution. (**D**) The result of the image by subtraction. The resolution is 2*(pixel size)/$k_c$, where $k_c$ is expressed in normalized frequencies.

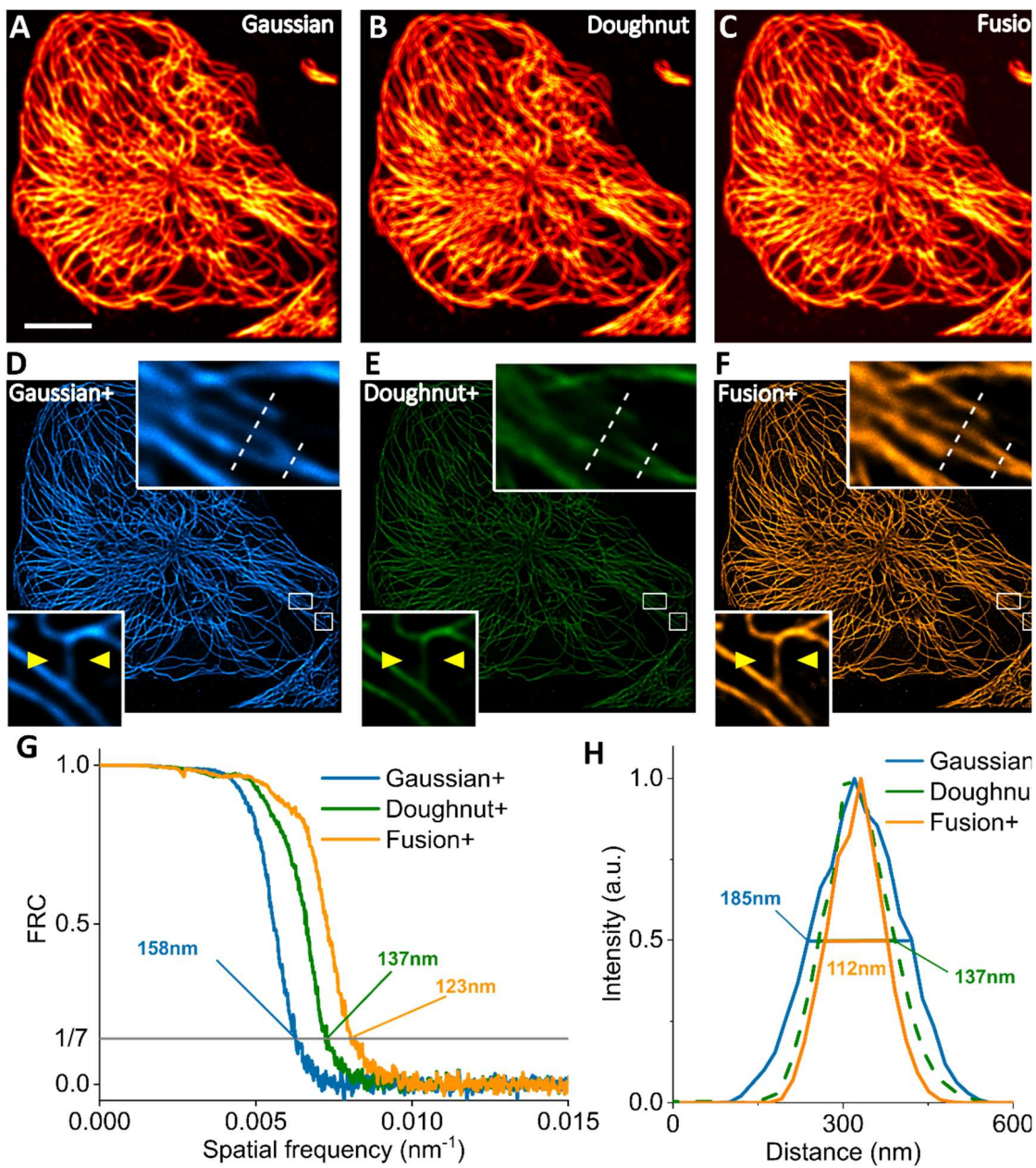

**Fig. S9. Simulation of Fourier domain heterochromatic fusion imaging of cell microtubules.** The structure is imaged by **(A)** Gaussian scanning, **(B)** Doughnut scanning and **(C)** Fourier domain heterochromatic fusion. The deconvoluted results with the same iteration are shown at **(D)**, **(E)** and **(F)**. Insets of (D–F) are the magnified regions of the microtubule. **(G)** The FRC resolution of (D-F). **(H)** Line profiles along the yellow arrow in (D-F). The scale bar is 10 μm.

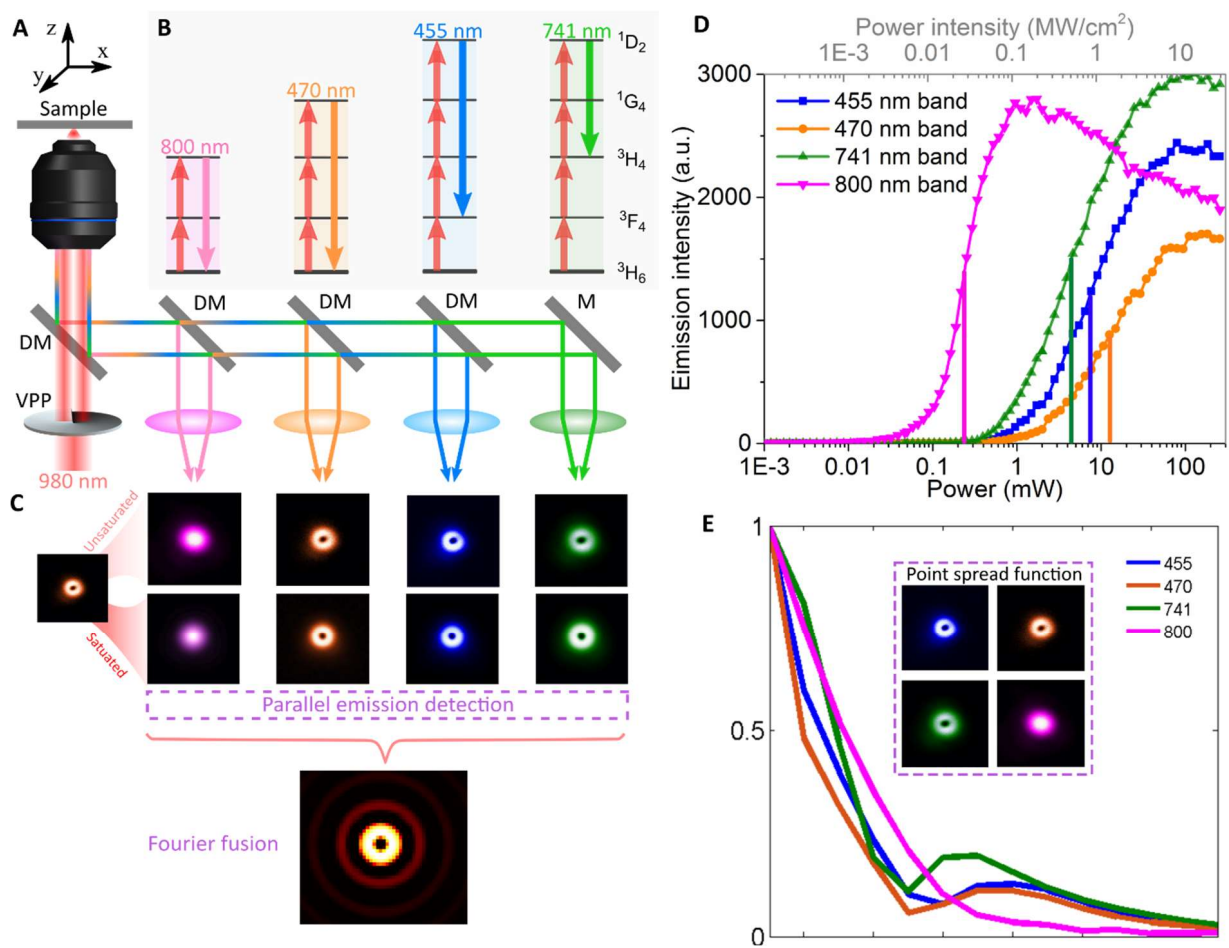

**Fig. S10. Schematics and optical properties of UCNPs for hyper-spectrum Fourier domain heterochromatic fusion in single beam scanning super-resolution microscopy.** (**A**) Optical setup for multi-colour detection under a single doughnut excitation beam. (**B**) The simplified energy level of UCNPs, 800 nm emission from two-photon level, 470 nm from three-photon level, 455 nm and 741 nm from four-photon level. (**C**) Different emission PSFs can be acquired from four distinct upconversion emission bands. The pattern can be modified by changing excitation power. The four-emission pattern can be applied to the Fourier domain fusion method. (**D**) The power dependence curve of the upconversion emissions from the four emission bands. (**E**) The optical transfer functions of the four emissions PSFs.

13